\documentclass[a4paper]{jpconf}  
\pdfoutput=1 
\usepackage{graphicx}  
\usepackage{dcolumn}   
\usepackage{bm}        
\usepackage{amssymb}   
\usepackage{comment}
\usepackage{amsmath}
\usepackage{amsfonts}
\usepackage{slashed}
\usepackage{graphicx}
\usepackage{float}
\usepackage{cancel}
\usepackage{hyperref}
\usepackage{color}
\newcommand{\beq}{\begin{equation}}
\newcommand{\eeq}{\end{equation}}
\newcommand{\beqa}{\begin{eqnarray}}
\newcommand{\eeqa}{\end{eqnarray}}
\newcommand{\bit}{\begin{itemize}}
\newcommand{\eit}{\end{itemize}}
\begin{document}

\title{$SO(10)$ Grand Unification from $M$ theory on a $G_2$ manifold}

\author{Miguel Crispim Rom\~{a}o$^1$}

\address{$^1$ School of Physics and Astronomy, University of Southampton, Southampton, SO17 1BJ, UK}

\ead{m.crispim-romao@soton.ac.uk}

\begin{abstract}
We consider Grand Unified Theories based on $SO(10)$ which originate from $M$ theory on $G_2$ manifolds. In this framework we are naturally led to a novel solution of the doublet-triplet splitting problem involving an extra $\overline{{\bf 16}}_X+{\bf 16}_X$ vector-like pair by considering discrete symmetries of the extra dimensions and preserving unification. Since Wilson line breaking preserves the rank of the gauge group, the necessary $U(1)$ gauge breaking is generated from extra multiplets. The main prediction of the approach is the existence of light states with the quantum numbers of a $\overline{{\bf 16}}_X+{\bf 16}_X$ vector-like pair which could show up in future LHC searches.
\end{abstract}

\section{Introduction}

The Large Hadron Collider (LHC) discovery of a Standard Model (SM) like Higgs \cite{Aad:2012tfa,Chatrchyan:2012ufa,CMS-PAS-HIG-14-009} stresses the need to understand the nature of the Electroweak Scale (EWS) and its stability. This accounts for solving the hierarchy, or naturalness, problem where the EWS is 16 orders of magnitude lower than the Planck scale for no apparent reason within the SM.

Supersymmetry (SUSY) offers an elegant framework where the mass parameters of scalars, and therefore the masses defining the EWS, are stabilised as their quantum corrections do not push them to Planckian values. Although stabilising the scalar masses, global SUSY does not provide an answer to what values they should take. In its local form, supergravity (SUGRA), the framework is extended in a way that mass  and SUSY breaking parameters can be generated through different mechanisms, although the specific values they take are dependent on the ultra-violet (UV) completion of the theory.

In order to accomplish a stabilisation in a \emph{natural} construction, SUSY models seem to require the masses of the new superparticles to be just above the EWS. Despite that, experimental bounds on superparticle masses suggest they have to be above $\mathcal{O}(1 \mbox{ TeV})$ \cite{Aad:2014lra,Chatrchyan:2013lya,Aad:2014nua}.

Despite the lack of experimental evidence for SUSY, we note that most of the hitherto SUSY phenomenology as been carried out in the context of the Minimal Supersymmetric Standard Model (MSSM) or special parts of its parameter space. Therefore, the existing experimental bounds constrain mostly the MSSM and leave other models -- more general and complete -- relatively untested. It is then necessary to study more general SUSY models and their limits. In principle there are infinite extensions and generalisations of the MSSM, and so we need a consistent and motivated guide when constructing such models.

String theory phenomenology has been a fruitful guide when constructing consistent supersymmetric models beyond the SM (BSM). In fact, string phenomenology has proven to have all the required ingredients when looking for a supersymmetric four dimensional theory with an UV completed quantum theory of gravity.

Nowadays we understand the different string theories (Type I, IIA, IIB, Heterotic-O, Heterotic-E) to be different limits of an underlying 11 dimensional theory, the so called $M$ theory, while being related to each other by a complex web of dualities. A final microscopic formulation of $M$ theory is yet to be developed, but the study of its low-energy limit -- believed to be the unique 11 dimensional SUGRA -- has yelled many advancements and insights about it. As such one is then led to study the phenomenology of this 11 dimensional supergravity upon suitable compactification of the extra seven dimensions. In fact, results over the past decade or so have shown that the simple combination of SUSY breaking and moduli stabilisation in string/$M$ theory can be a very useful guide to constructing models \cite{Acharya:2007rc,Acharya:2008zi,Acharya:2012tw,Kachru:2003aw,Balasubramanian:2005zx}. The framework has been shown to lead to effective supersymmetric models with distinctive features and very few parameters.

In $M$ theory we are led to consider supersymmetric grand unified theories (SUSY GUTs) based on simple groups, such as $SU(5)$, which explain the fermion quantum numbers and unify the three SM forces. Such models are usually phenomenologically rich and furnish potentially realistic low energy theories. We have to face many of the usual problems of this type of theories, namely the basic problem of GUTs: doublet-triplet splitting problem. In GUTs, the SM Higgs doublet is unified into a GUT multiplet containing colour triplets which can mediate proton decay too quickly as, in virtue of the GUT symmetry, the doublet and the triplet should have the same EWS mass. The most common solution to this problem present in many models, including those originating in string/$M$ theory framework, is to make the colour triplets very massive \cite{Witten:1985xc,Mohapatra:2007vd,Lee:2010gv} as to suppress the proton decay channel. This is often achieved with a discrete symmetry whose effective action on the triplets is different from that on the doublets, i.e. a symmetry that does not commute with the GUT symmetry.

It is well known that $M$ theory on a manifold of $G_2$ holonomy leads elegantly to four dimensional models with $N=1$ supersymmetry. In this framework, both Yang-Mills fields and chiral fermions arise from very particular kind of singularities in the extra dimensions \cite{Acharya:2001gy,Acharya:2004qe}. Yang-Mills fields are localised along three-dimensional subspaces of the seven extra dimensions along which there is an orbifold singularity. Chiral fermions, coupling to the Yang-Mills fields, arise from additional localised points at which there is an $A,D,E$-type conical singularity. Therefore, different GUT multiplets are
localised at different points in the extra dimensions. 
The GUT gauge group can be broken to $SU(3)\times SU(2) \times U(1)$ (possibly with additional $U(1)$ factors) by Wilson lines on the three-dimensional subspace supporting the gauge fields.
Compact manifolds of $G_2$ holonomy -- being Ricci flat and having a finite fundamental group -- do not admit continuous symmetries, but could be endowed with discrete symmetries. If present, such symmetries play a very important role in the physics. In particular, for the $SU(5)$ case, Witten showed that such symmetries can solve the doublet-triplet splitting problem \cite{Witten:2001bf}. 

In this work we will extend the scope of the $M$ theory approach from the previously considered $SU(5)$/MSSM case arising from $M$ theory on $G_2$ manifolds -- the so-called $G_2$MSSM -- to $SO(10)$ \cite{Acharya:2015oea}, where an entire fermion family  $Q,u^c,d^c,L,e^c,N$, including a charge conjugated right-handed neutrino $N$, is unified within a single $\mathbf{16}^m $ representation. In particular we focus on the Higgs doublet-triplet splitting problem and present a solution, which turns out to be necessarily quite different from the $G_2$-MSSM. As a consequence, our solution leads to distinct phenomenological constraints and predictions.

In the remainder of this work, we start by reviewing some basic ideas and results from $M$ theory within the example of 
$SU(5)$/$G_2$-MSSM. This will allow for a introduction on the main features of $M$ theory model building with the phenomenological features being explicitly discussed. After we discuss the $SU(5)$ case, we will then embark on a discussion of the new $SO(10)$ case and how we are led to additional light states at the reach of the LHC.

\section{The $G_2$-MSSM, an $SU(5)$ review}

We start by reviewing some key ideas and features of model building in $M$ theory, and their application in the $SU(5)$ context with a low-energy spectrum similar to the MSSM, the so called $G_2$-MSSM \cite{Acharya:2008zi}.

Just like in most string theory frameworks, in $M$ theory the compactified space plays a crucial role  in defining the effective field theory of the 4 dimensional spacetime we live in. Namely, the matter fields of the 4 dimensional theory are localised in the internal space close to singularities, while gauge fields are supported by 3 dimensional subspaces that support the matter fields' singularities. Furthermore, the topology of the internal space can allow for the GUT group to be broken by Wilson line phases that arise when the compactified space has a non-trivial fundamental group, i.e. when it has \emph{holes} or \emph{handles}. To see this, consider that the compactified space has non-contractible closed loops (or 1-cycles), then there will be non-vanishing gauge field configurations such that, along a non-contractible loop, we have
\beq
\mathcal{W} = \mathcal{P} \tr \exp \oint A_k d^k \neq 1,
\eeq
where $A$ is the GUT gauge field, $\tr$ is a group trace due to the non-abelian nature of the GUT group, $\mathcal{P}$ represents path-ordering, and $k=5,...,11$ runs through the internal space dimensions.

The quantities $\mathcal{W}$ cannot be gauged away as they are not local gauge quantities, and as such are observables. They can, nonetheless, be absorbed by a matter field localised in a singularity, which is itself along a Wilson line. These will break the GUT group into a smaller group, which will be formed by all the elements of the GUT group that commute with $\mathcal W$. To see this consider a singularity in the extra dimensions supporting a GUT irrep $\Psi$, which is localised along a Wilson line, and let it absorb the respective quantities $\mathcal W$
\beq
\Psi \to \Psi^w = (\mathcal{W}\Psi),
\eeq
the surviving gauge group will now be composed of the elements of the GUT group, $g$, that can commute with, i.e. ``pass through'', $\mathcal W$
\beq
\Psi^w \to (\mathcal{W} g \Psi) = g \Psi^w.
\eeq

It is also crucial to notice that $\mathcal W$ furnishes a representation of the fundamental group as the integral of closed loops is itself a homotopy quantity. Hence, if the fundamental group is taken to be the abelian $\bf{ Z_n }$, then $\mathcal{W}^n=1$.

If we take $\bf{ Z_n }$ to be the fundamental group, then all the Wilson lines will commute with each other. As such they will be elements of the surviving gauge group. Being commuting elements of the surviving group, they will be then embedded in its centre. This means that the Wilson line matrices are then generated by the generators of the surviving $U(1)$ groups. We can then represent a Wilson line by

\beq
\mathcal W = \sum_m \frac{1}{m!} \left( \frac{i 2 \pi }{n} \sum_j a_j Q_j  \right)^m \ ,
\label{eq:Wrep}
\eeq
which is just a convenient embedding of the Wilson line in the surviving group, in analogy with the usual representation of a group element
\beq
g = \sum_m \frac{1}{m!} \left( i \sum_j \epsilon_j Q_j \right)^m \ ,
\eeq
taking\footnote{With some abuse of notation, as $\epsilon_i$ are local functions while $a_i$ are constants.} $\epsilon_i\to a_i 2 \pi/n$ such that it explicitly holds $\mathcal W^n=1$, and $Q_i$ are the generators of the $U(1)$ factors of the surviving gauge group. This also means that the Wilson line is a diagonal element of the GUT group, and hence the symmetry breaking patterns allowed by the Wilson line will be rank preserving, i.e. it allows for the same breaking patterns as an adjoint valued Higgs.

The Wilson line phases play a crucial role for model building. In the context of $SU(5)$, Witten \cite{Witten:2001bf} notice that if the geometry of the compactified space admits a geometrical symmetry isomorphic, i.e. of the same type, than the fundamental group -- here taken to be ${\bf Z_n}$ -- then the action of the discrete symmetry will be altered by the Wilson line phases when acting on the matter fields that absorbed them. This means that the Wilson line phases will effectively act as charges of the discrete symmetry and therefore the resulting field theory will have a \emph{naturally arising} discrete symmetry that does not commute with the initial GUT group. We can then use these charges to further constrain the superpotential couplings and look for an $M$ theory solution with realistic low energy Lagrangian.

The allowed tree-level superpotential terms, i.e. those allowed by gauge symmetry and by the discrete symmetry, will have coefficients given by the action of membrane instantons on the three dimensional space supporting the fields in the compactified extra dimensions. The action of the instantons can be parametrised by the distances between the singularities supporting the relevant fields. For example, take a trillinear superpotential coupling between three supermultiplets $X$, $Y$,and $Z$, we have then
\beq
y X Y Z : \ y\sim \exp \left( - \mbox{vol}_{XYZ}\right) \ ,
\eeq
while the billinear terms, i.e. $\mu$-type mass parameters, we will have suppressions of a GUT scale mass parameter which can take either small or large values depending on the distance between the positions the multiplets have inside the compactified space. Furthermore, one can find that the unified gauge coupling is parametrised by the volume of the seven dimensional internal space, $V_7$, 
\beq
V_7 \sim \frac{1}{\alpha_{GUT}^{7/3}}\ .
\eeq

Ultimately, the specific values for the above quantities can only be computed if one has a full geometric description of the internal space. While such is yet and open problem, the above sketch serves as a general guideline to what type of values we should expect.

As an illustrative example, we will now see in some detail how these ideas are used such that Witten's $M$ theory approach to $SU(5)$, the combination of the discrete symmetry, the Wilson lines, and the fact that GUT multiplets are localised at points, allows us to have a natural solution for the doublet-triplet problem.

Following the representation for a Wilson line from equation \ref{eq:Wrep}, for the $SU(5)$ GUT the Wilson line will be generated by the Hypercharge generator. Under a choice of normalisation for the coefficient in equation  \ref{eq:Wrep}, the Wilson line for $SU(5)$ can be written as
\beq
\mathcal{W} = \mbox{diag} \left( \eta ^\gamma, \eta ^\gamma, \eta ^\gamma, \eta ^\delta, \eta ^\delta \right) ,
\eeq
where $\eta \equiv e^{2\pi i /n}$, and $2\delta + 3\gamma = 0$ mod $n$ .

Consider that the compactified space has the appropriate singularities to support the MSSM spectrum and that a GUT multiplet containing one of the MSSM Higgses absorbs Wilson line phases. In more detail, taking the geometric discrete symmetry to be ${\bf Z_n}$, if $\overline{D}^w, H_d^w \in \overline{\mathbf{5}}^w $  is localised along the Wilson line, $D^h, H^h_u \in {\bf 5}^h$ the multiplet containing the second MSSM Higgs, and ${\bf {\overline 5}}^m,{\bf 10}^m$ the matter multiplets, under the symmetry the GUT multiplets will transform as
\begin{align}
\overline{\mathbf{5}}^w &\to  \eta^{\omega} \left( \eta^{\delta} H^w_d \oplus \eta^{\gamma} \overline{D}^w \right) ,\nonumber \\
\mathbf{5}^h &\to \eta^{\chi} \mathbf{5}^h , \label{eq:w5} \\
\overline{\mathbf{5}}^m &\to \eta^{\tau} \overline{\mathbf{5}}^m , \nonumber \\
\mathbf{10}^m &\to \eta^{\sigma} \mathbf{10}^m , \nonumber
\end{align}
where $\omega$, $\chi$, $\tau$, and $\sigma$ are overall charges of the discrete symmetry.

In order to be potentially realistic, a low-energy model resembling the MSSM requires the constraints on the discrete charges to allow for Yukawa couplings, Majorana neutrino masses, and colour-triplet masses must be present. The required couplings and the respective discrete charge constraints are presented in table \ref{tab:g2mssm}, where we take $\omega = 0$ without loss of generality.

\begin{table}[h]
\caption{\label{tab:g2mssm}  Couplings and charges for $SU(5)$ operators.}
\begin{center}
\begin{tabular}{ll}
\br
Coupling & Constraint   \\ 
\mr
$H_u^h \mathbf{10}^m \mathbf{10}^m$ & $2 \sigma + \chi = 0$ mod $n$ \\ 
$H_d^w \mathbf{10}^m \overline{\mathbf{5}}^m$ & $\sigma + \tau +\delta = 0$ mod $n$ \\ 
$ H_d^w H_d^w  \overline{\mathbf{5}}^m  \overline{\mathbf{5}}^m$ & $2\chi + 2\tau = 0$ mod $n$ \\ 
$\overline{D}^w D^h$ & $\chi + \gamma = 0$ mod $n$\\ 
\br
\end{tabular}
\end{center}
\end{table}

One can solve these by writing all charges in terms of, say, $\sigma$
\begin{align}
\chi &= -\gamma = -2 \sigma \mod n, \nonumber \\
\delta &= -3\sigma + n/2 \mod n , \\
\tau &= 2\sigma + n/2 \mod n, \nonumber
\end{align}
which {\it automatically forbids the $\mu$-term and dimension four and five proton decay operators}, provided that the respective discrete charges balances forbid the respective couplings.

As a result, the discrete symmetry forces a tree-level $\mu=0$ while allowing for a high mass for the colour-triplets assuming the singularities supporting $\overline{\bf 5}^w$, ${\bf 5}^h$ are close to each other. However,
phenomenologically and experimentally we know $\mu \geq$ 100 GeV, from direct limits on the masses of charged Higgsinos in colliders. The symmetry must therefore be broken at low-energies, while fully effective at high-energies. 
In $M$ theory all the moduli are geometric, and hence they are naturally charged under the discrete symmetry. As it was shown in \cite{Acharya:2008hi} -- for compactifications of $M$ theory on a $G_2$ manifold -- strong gauge dynamics in the hidden sector generate contributions to the moduli potential which can be minimised while fixing all the moduli. In this process, supersymmetry is spontaneously broken in the visible sector at a hierarchically small scale.

Furthermore, as the moduli are charged under the discrete symmetry, generically the vacua of the potential will spontaneously break the ${\bf Z_n}$ symmetry. This then generates an effective $\mu$ term from Kahler potential operators involving both moduli and visible sector fields, of the form
\beq
K \supset \frac{s}{m_{pl}} H_u H_d \;\;\;+\;\;h.c.,
\eeq
similar to the Giudice-Masiero mechanism \cite{Giudice:1988yz}, where $s$ generically denotes a modulus field and $m_{pl}$ is the Planck scale. 
From previous studies on moduli stabilisation \cite{Acharya:2007rc,Acharya:2008zi,Acharya:2012tw,Acharya:2008hi}, we know that the moduli vevs are stabilised approximately at $\langle s \rangle \sim 0.1 m_{pl}$,  with SUSY breaking non-vanishing $F$-terms vevs being $\langle F_s \rangle \sim m_{1/2} m_{pl} $ where $m_{1/2}$ is the gaugino mass. From the standard supergravity Lagrangian \cite{Brignole:1997dp}, one can find that the $\mu$ term can be generated by the above Kahler potential interactions through
\beq
\mu = \langle m_{3/2} K_{H_u H_d} - F^k K_{H_u H_d k} \label{eq:muterms}\rangle ,
\eeq 
which leads to
\beq
\mu \sim \frac{\langle s \rangle}{m_{pl}} m_{3/2} + \frac{\langle F_{s} \rangle }{m_{pl}} .
\eeq

In $M$ theory the gaugino masses are suppressed \cite{Acharya:2007rc,Acharya:2008zi,Acharya:2012tw,Acharya:2008hi} relatively to the gravitino mass, $m_{3/2}$, which in this framework can be estimated to be of order $\mathcal{O}(10\mbox{ TeV})$. Therefore, the $F$-term vev is subleading and we estimate the $\mu$ term to take a naturally desirable phenomenological value
\beq
\mu \sim 0.1 m_{3/2} \sim \mathcal{O}(\mbox{TeV}) \ .
\eeq

Moduli stabilisation and consequent vevs will spontaneously break SUSY in the visible sector with universal soft-masses being set by the value of $m_{3/2 }\sim \mathcal{O}(10 \mbox{ TeV})$. Therefore, we expect lighter supersymmetric fermionic partners (gauginos, Higgsinos) than scalar partners, as both $\mu$ and $m_{1/2}$ mass parameters are naturally suppressed relative to $m_{3/2}$.

\section{$SO(10)$ GUT models in $M$ theory}

Having reviewed the key features and ingredients for model building in $M$ theory with the $SU(5)$ GUT group, we now develop our $M$ theory approach to $SO(10)$ \cite{Acharya:2015oea}. We will see that we cannot implement the same doublet-triplet solution as above, and hence we are forced to consider a new solution with very distinct phenomenological implications.

As the Wilson line breaking mechanism is rank preserving, it can break the $SO(10)$ down to 
\beq
SU(3)\times SU(2)\times U(1)_Y \times U(1) ,
\eeq
which means that $\mathcal W$ phases are linear combinations of two independent $U(1)$ factors, and as such is fully parametrised by two independent phases.

As for the case of the $G_2$-MSSM described in the previous section, we assume the internal geometry to have a  ${\bf Z_n}$ symmetry. In analogy to equation \ref{eq:w5}, and after a choice of normalisation for the coefficients in equation \ref{eq:Wrep}, a fundamental of $SO(10)$, ${\bf 10}$, that absorbs Wilson line phases can be found to transform as
\begin{equation}\label{eq:W10}
\mathbf{10}^w \to \eta^{\omega} \left(\eta^{-\alpha} H_d^w  \oplus \eta^{\beta} \overline{D}^w 
\oplus  \eta^{\alpha} H_u^w \oplus  \eta^{-\beta} D^w
\right).
\end{equation}

If we take the minimal $SO(10)$ spectrum, the $\mu$-term arises from a GUT term in the superpotential of the form 
\beq
W \supset \mu \mathbf{10}^w\mathbf{10}^w = \mu \left(H^w_u H^w_d + D^w  \overline{D}^w\right) ,
\eeq
and as such we come across the doublet-triplet spliting problem as the triplet and the Higgses share the same value for their mass, $ m_D= \mu$. Clearly if $2\omega = 0 \mod n $ both terms are allowed, otherwise both terms are forbidden.
{\it Therefore generically, they will both be forbidden.} We conclude that the same solution for the doublet-triplet spliting problem used for the $SU(5)$ case cannot be employed in the $SO(10)$ case.

While looking for a spectrum without light extra colour triplets, we contemplate the possibility of extra ${\bf 10}$ multiplets whose bilinear couplings with ${\bf 10}^w$ would allow for only one light pair of Higgses. With additional ${\bf 10}$ multiplets, we can allow or forbid distinct couplings between the different states of the various ${\bf 10}$ multiplets. However, in the end we find that there will typically be more than one pair of light Higgs doublets, which tend to destroy gauge coupling unification. To see this explicitly, consider one additional ${\bf 10}$, denoted $\mathbf{10}^h$, without Wilson line phases, i.e. transforming under the discrete symmetry as $\mathbf{10}^h \rightarrow \eta^\xi \mathbf{10}^h$. In total, there are eight possible gauge invariant couplings with a $\mathbf{10}^w$ and $\mathbf{10}^h$ that can be written in matrix form as
\begin{align}
W \supset \mathbf{H}_d^T \cdot \mu_H \cdot \mathbf{H}_u + \overline{\mathbf{D}}^T \cdot M_D \cdot \mathbf{D} ,
\end{align}
where $\mu_H$ and $M_D$ are $2\times 2$ superpotential mass parameters matrices, $\mathbf{H}_{u,d}^T = \left(H^w_{u,d}, H^h_{u,d}\right)$, $\mathbf{\overline{D}}^T = \left (\overline{D}^w,\overline{D}^h\right)$, and $\mathbf{D}^T = \left(D^w,D^h\right)$. The tree-level entries of the matrices are non-vanishing or vanishing depending on which of the following discrete charge combinations are zero (mod $n$) or not, respectively
\begin{align}
D^w\overline D^w , \ H^w_u H^w_d & : \  2 \omega   , \nonumber \\
D^h\overline D^h , \  H^h_u  H^h_d & : \ 2 \xi   ,\nonumber \\
H^w_u  H^h_d &: \ \alpha + \omega + \xi   , \label{eq:10w10m} \\
H^h_u  H^w_d &: \ -\alpha +\omega+\xi   , \nonumber\\
D^w \overline D^h &: \ -\beta +\omega + \xi    , \nonumber\\
D^h \overline D^w &: \ \beta +\omega +\xi  .\nonumber
\end{align}
The naive doublet-triplet splitting solution would be for $\mu_H$ to have only one vanishing eigenvalue, with $M_D$ having all non-zero eigenvalues. One finds that there is no choice of constraints in equation \ref{eq:10w10m} that accomplishes this scenario. Henceforth, we shall only consider a single light $\mathbf{10}^w $, without any extra ${\bf 10}$ multiplets at low energies.

It then seems to be necessary a new solution for the doublet-triplet splitting problem. Namely, we need to prevent fast proton decay channels that can be generated by the presence of light colour triplets scalars; and to re-establish gauge coupling unification, which is spoiled by the presence of extra MSSM light states that do not form a complete GUT multiplet.

In order to prevent fast proton decay processes, assuming a single light $\mathbf{10}^w $, we note that it is possible to use the discrete symmetry to forbid certain couplings, 
namely to {\it decouple $D^w$ and $\overline{D}^w$ from matter}.
Such couplings arise from the operator
\beq
\mathbf{10}^w \mathbf{16}^m \mathbf{16}^m,
\eeq
 with  $\mathbf{16}^m$ denoting the three $SO(10)$ multiplets, each containing a SM family plus right handed neutrino $N$.
Letting $\mathbf{16}^m$ transform as $\eta^m \mathbf{16}^m$, the couplings and respective charge constraints are shown table \ref{tab:SO10}, where we allow for up-type quark and right-handed neutrino Yukawa couplings,
\beq
y_u^{ij}H_u^w \mathbf{16}^m_i \mathbf{16}^m_j \equiv y_u^{ij}H_u^w (Q_iu_j^c+L_iN_j +i\leftrightarrow j),
\label{yu}
\eeq
and similarly for down-type quark and charged leptons.

\begin{table}[h]
\caption{\label{tab:SO10}  Couplings and charges for $SO(10)$ operators.}
\begin{center}
\begin{tabular}{ll}
\br
Coupling & Constraint \\ 
\mr
$H_u^w \mathbf{16}^m \mathbf{16}^m$ & $2m + \alpha + \omega = 0$ mod $n$ \\ 
$H_d^w \mathbf{16}^m \mathbf{16}^m$ & $2m - \alpha + \omega = 0$ mod $n$ \\ 
$D^w \mathbf{16}^m \mathbf{16}^m$ & $2m - \beta + \omega \neq 0$ mod $n$ \\
$\overline{D}^w \mathbf{16}^m \mathbf{16}^m$ & $2m + \beta + \omega \neq 0$ mod $n$\\ 
\br
\end{tabular}
\end{center}
\end{table}

A solution for the these constraints can be found and leads to a tree-level superpotential that allows for matter Yukawas and decoupled light colour triplets. The later feature has been previously considered by Dvali in \cite{Dvali:1995hp} and \cite{Kilian:2006hh,Reuter:2007eh,Howl:2007zi} from a bottom-up perspective. 

As it was argued above, $M$ theory on a $G_2$ manifold vacua breaks the discrete symmetry. Hence, we expect that proton-decay operators forbidden by the discrete symmetry to be effectively generated as the moduli are stabilised and acquire a vev. The relevant operators can be generated in the Kahler potential, schematically, writing $D=D^w$,
\beq
K \supset \frac{s}{m_{pl}^2} D Q Q +\frac{s}{m_{pl}^2} D e^c u^c + \frac{s}{m_{pl}^2} D N d^c  + \frac{s}{m_{pl}^2}\overline{D} d^c u^c + \frac{s}{m_{pl}^2}\overline{D} Q L .
\eeq
In a similar procedure to the one that generates effective $\mu$-terms from equation \ref{eq:muterms}, the effective superpotential may be calculated from supergravity to be
\beq  
W_{eff}  \supset \lambda D Q Q  + \lambda D e^c u^c + \lambda D N d^c +\lambda \overline{D} d^c u^c+\lambda \overline{D} Q L,
\label{eq:proton}
\eeq
where the coefficients $\lambda$ can be estimated
\beq
\lambda \approx \frac{1}{m^2_{pl}}\left( \langle s \rangle m_{3/2} + \langle F_{s}\rangle  \right) \sim 10^{-14} .
\eeq

Unlike the case of $SU(5)$, there is no $SO(10)$ invariant bilinear term $\kappa L H_u$, where in our framework $\kappa \sim \mathcal{O} (1\mbox{ TeV})$, which would lead to fast proton decay. The terms in equation (\ref{eq:proton}) allow for colour triplet scalar induced proton decay, which can be estimated by
\beq
\Gamma_p \approx \frac{\left| \lambda^2 \right|^2}{16 \pi^2}\frac{m_p^5}{m_D^4} .
\eeq
In general we expect the mass of the colour triplets to be of the same order as $\mu$, i.e., $m_D \sim 10^3$ GeV and as such the proton lifetime is
\beq
\tau_p= \Gamma_p^{-1} \sim 10^{38} \;\mbox{yrs} ,
\eeq
which exceeds the current experimental limit.

It is also worthwhile to estimate the lifetime of the colour triplets as they can spoil BBN. Taking an approximation where the products of the decay are massless, the lifetime
\beq
\tau_D  = \Gamma_D^{-1} \sim \left( \lambda^2 m_D\right)^{-1} \sim 0.1 \;\mbox{sec} .
\eeq
is short enough to avoid the constraints from BBN, while long enough to give interesting collider signatures.

Gauge coupling unification is in general spoiled by the presence of light states beyond the MSSM spectrum that do not belong to a complete GUT irrep. In our case, gauge coupling unification is spoiled by the presence of the light colour triplets from ${\bf 10}^w$ containing the MSSM Higgses.

The immediate (and the only one that we know) solution to this problem is then to complete the colour triplets into complete GUT multiplets. In order to accomplish this, we need to make use of the discrete symmetry to split some extra GUT multiplets such that we can integrate out -- i.e. make very heavy -- states that have the same SM quantum numbers as the colour triplets. As we have shown that we cannot split a ${\bf 10}$ in the desired way, we then consider the presence of an extra vector-like pair of ${\bf 16}$'s, labelled ${\bf 16}_X+{\bf {\overline{16}}}_X$, since they contain the states  $d^c_X,\overline{d^c}_X$ carrying the same SM quantum numbers as $D^w$, $\overline{D}^w$. Under the assumption that the singularities supporting ${\bf 16}_X+{\bf {\overline{16}}}_X$ are close enough, the discrete symmetry can then allow for a membrane instanton generated GUT-scale mass for $d^c_X,\overline{d^c}_X$ by allowing this bilinear coupling at tree-level, while keeping the remaining states light, by forbidding their presence at tree-level.

To see this more explicitly, take ${\bf 16}_X$ to be localised along a Wilson line. In a similar procedure that leads to equation \ref{eq:W10}, one can find that ${\bf 16}_X$ will now transform under the discrete symmetry as
\beq
{\bf 16}_X \to  \eta^x \left( \eta^{-3\gamma} L \oplus \eta^{ 3\gamma+\delta} e^c \oplus  \eta^{3 \gamma - \delta} N \oplus  \eta^{-\gamma-\delta} u^c  \oplus \  \eta^{-\gamma +\delta} d^c \oplus  \eta^{\gamma} Q \right) .
\eeq
Furthermore, assuming ${\bf {\overline{16}}}_X$ transforms without Wilson line phases, i.e. ${\bf {\overline{16}}}_X \to \eta^{\overline x}\,  {\bf {\overline{16}}}_X$, the condition for the presence of a GUT-scale mass term is 
\begin{equation}
\overline{d^c}_X d^c_X : x - \gamma + \delta + \overline{x} = 0 \mod n ,\ 
\end{equation}
while forbidding all the other couplings that would arise from ${\bf 16}_X {\bf {\overline{16}}}_X$.

\begin{figure}
\begin{center}
\includegraphics[scale=1]{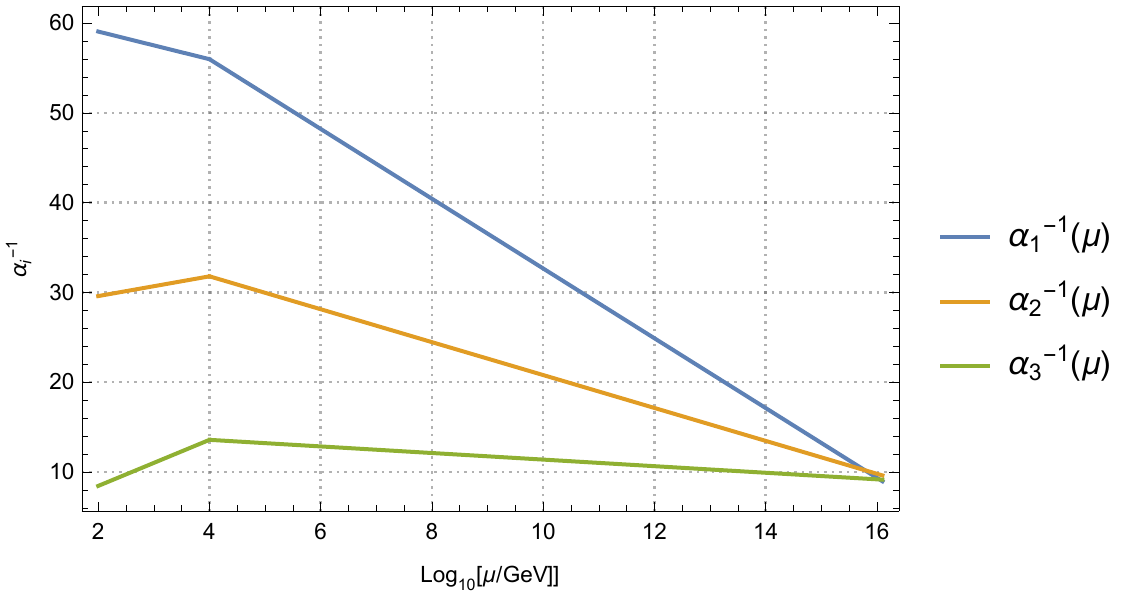} 
\caption{Gauge coupling unification for $SO(10)$ with extra ${\bf 16}_X+{\bf {\overline{16}}}_X$}\label{fig:gaugecouplings}
\end{center}
\end{figure}

The light $D^w$, $\overline{D}^w$ from ${\bf 10}^w$ will then ``complete" the 
${\bf 16}_X+{\bf {\overline{16}}}_X$ pair. The remaining, light, states of ${\bf 16}_X+{\bf \overline{16}_X}$ will
obtain $\mu$-term type masses via the Kahler potential of order a TeV, via the Giudice-Masiero mechanism of equation \ref{eq:muterms}.
As we can see in figure \ref{fig:gaugecouplings} gauge unification is restored, with a larger gauge coupling
at the GUT scale due to the extra low energy contributions to the MSSM spectrum.

One can worry if the new vector-like family will induce dangerous mixing with the quarks and leptons of the MSSM as in general the moduli vevs will induce $\mu$-terms of the form $\mu \overline{\bf 16}_{X}{\bf 16}^m $. But as only $\overline{\bf 16}_{X}$, and not $\mathbf{16}_X$, couples to matter through the effective $\mu$-terms, we find that all the light components of extra matter decouple from ordinary matter . As an illustrative example we consider the up-type quark sector. The (effective) superpotential contribution to the mass matrix is, schematically, $W \supset \overline{U}^T \cdot  M_U \cdot  U$, where $U^T = \left(u_i , \overline{u^c}_X , u_X\right)$, $\overline{U}^T = \left(u^c_i, \overline u_X , u^c_X\right)$, with $i=1,2,3$, and 
\begin{equation}
M_U \sim 
\begin{pmatrix}
y_u^{ij} \langle H_u \rangle & \mu^i_{X} & \lambda^i_{X}\langle H_u \rangle \\
\mu^j_{X} &  \lambda_{XX} \langle H_u \rangle & \mu_{XX} \\
\lambda^j_{X } \langle H_u \rangle &\mu_{XX} & \lambda_{XX} \langle H_u \rangle
\end{pmatrix}.
\end{equation}
Here $\mu^i_{X},\mu^j_{X},\mu_{XX}$ are moduli induced $\mu$-type mass parameters of $\mathcal{O}(\mbox{TeV})$ while $\lambda^i_{X},\lambda^j_{X},\lambda_{XX}$ are moduli induced trilinear interactions
which are vanishingly small, $\mathcal{O}(10^{-14})$. As a result, the 
determinant of $M_U$ is approximately independent of $\mu^i_{X},\mu^j_{X}$, i.e. of the mixing masses, to leading order in $\lambda$'s. This means the eigenvalues are, to leading order in $\lambda$'s, independent of the mixing so the up-type quarks decouple from the new particles. The same study for other sectors leads to the same conclusion.

The inclusion of the extra vector-like family comes with other benefits for model building. As the Wilson line symmetry breaking mechanism is rank preserving, the extra $U(1)$ has to be spontaneously broken at the field theory level as the presence of a massless $Z'$ is in clear contradiction with experiment. To accomplish this, and assuming D-flat direction, the ${\bf 16}_X+{\bf {\overline{16}}}_X$ acquire vevs in their right-handed neutrino components, $\langle N_X \rangle = \langle  \overline{N}_X \rangle =v_X$, breaking the rank. 
Here we do not specify the details of this symmetry breaking mechanism. However the scale $v_X$ is constrained, as discussed below.

Another consequence of a non-vanishing vev $v_X$ is the possibility to generate matter right-handed neutrino Majorana masses, which in turn would lead to light physical neutrino masses through a see-saw mechanism. This turns out to be fortunate as there are no $SO(10)$ irreps larger than $\mathbf{45}$ in $M$ theory. In the present framework, a Majorana mass term for a matter right handed neutrino is generated by letting the discrete symmetry to allow the Planck suppressed operator 
\beq
\frac{1}{m_{pl}}\overline{\bf 16}_{X}\overline{\bf 16}_{X}\mathbf{16}^m\mathbf{16}^m ,
\eeq
which  is allowed if the discrete symmetry charges satisfy $2 \overline{x}+2m=0\mod n$, and leads to the Majorana mass $M \sim \frac{ {v_X}^2}{m_{pl}}$.

Unlike in the $G_2$-MSSM, neutrinos will have the same Yukawa coupling as the up-type quarks $y_u^{ij}$, as in equation \ref{yu}, in virtue of the GUT group. This means that their Dirac masses are the same as the up-quark masses. For the case of the top quark mass we would need $M \sim 10^{14}$ GeV in order to give a realistic neutrino physical mass. This can only be achieved by the above see-saw mechanism if the rank breaking vev is $v_X \gtrsim 10^{16}$ GeV.

While the previous constraint offers a lower bound for $v_X$, its magnitude will also have an upper bound due to R-parity violating (RPV) dynamically generated operators, due to moduli and $N_X, \overline{N}_X$ vevs, arising from the Kahler interactions
\beq
K_{RPV} \supset  \frac{s}{m^3_{pl}} \mathbf{16}_X \mathbf{16}^m \mathbf{16}^m \mathbf{16}^m +  \frac{s}{m^2_{pl}} \mathbf{10}^w \mathbf{16}_X \mathbf{16}^m .
\eeq
As $s$ and $N_X$ acquire vevs, these operators generate the effective superpotential terms, which are otherwise forbidden by the discrete symmetry,
\beq
W^{eff}_{RPV} \supset \lambda \frac{v_X}{m_{pl}} L L e^c + \lambda\frac{v_X}{m_{pl}} LQ d^c +  \lambda \frac{v_X}{m_{pl}} u^c d^c d^c + \lambda v_X L H_u ,  \label{eq:RPV}
\eeq
with $\lambda \sim \mathcal{O}(10^{-14})$.

One can rotate away the last term into $\mu H_dH_u$ by a small rotation $\mathcal{O}(v_X/m_{pl})$ in $(H_d,L)$ space\footnote{Where we abuse notation and use the same letters for the rotated fields.}
\beq
W^{eff}_{RPV} \supset  y_e  \frac{v_X}{m_{pl}} L L e^c +  y_d   \frac{v_X}{m_{pl}} LQ d^c +  \lambda  \frac{v_X}{m_{pl}} u^c d^c d^c ,  \label{eq:RPVrot}
\eeq
where the first two terms originate from the Yukawa couplings $y_eH_dLe^c$, etc., and we have dropped their $\mathcal{O}(\lambda)$ contributions as  Yukawa rotated contributions are much larger. The last term is of $\mathcal{O}(\lambda)$ as it does not have new contributions from the Yukawa couplings. 

Although it was rotated away, the last term in equation \ref{eq:RPV} gives an important RPV limit that constrains the value of $v_X$ \cite{Acharya:2014vha,Banks:1995by,Hirsch:2000ef}, as it induces an extra contribution to the neutrino masses after the rotation.
 Neutrino masses limit the bilinear contribution to the mixing to be $ \lambda v_X \lesssim \mathcal{O}(1 \mbox{ GeV})$, which leads to the upper bound $v_X \lesssim 10^{14}$ GeV in contradiction with the see-saw requirement $v_X \sim 10^{16}$ GeV assumed in the above estimates.

In order to accommodate both bounds we need to suppress RPV induced neutrino masses while maintaining 
$v_X \sim 10^{16}$ GeV. This is possible if there is some suppression in the last coupling in equation \ref{eq:RPV} of order $\mathcal{O}(10^{-2})$. Such a suppression could occur if the discrete symmetry does not allow for this term, i.e. there is no modulus with appropriate charge coupling with ${\bf10}^w {\bf 16}_X 1{\bf 6}^m$ in the Kahler potential. However higher dimension operators with greater inverse powers of the Planck mass would still be expected providing the necessary extra suppression in order to relax the bound. Even though such accidental suppressions might be non-generic, there is no reason for them not to be present in examples.

The RPV terms in equation \ref{eq:RPVrot} are responsible for inducing the lightest supersymetric particle (LSP) decay. We can estimate its lifetime as \cite{Acharya:2011te}:
\begin{equation}
\tau_{LSP} \simeq \frac{10^{-13} \sec}{ \left( v_X/m_{pl}\right)^2}\left( \frac{m_0}{10 \ \mbox{TeV}}\right)^4 \left( \frac{100 \ \mbox{GeV}}{m_{LSP}}\right)^5 .
\end{equation}
Since, as discussed above, $v_X/m_{pl}\sim 10^{-2}$ for realistic neutrino masses, 
one finds $\tau_{LSP} \sim 10^{-9} \sec$. This value is compatible with current bounds $\tau_{LSP} \lesssim 1$ sec \cite{Dreiner:1997uz}, from BBN, but rules out the LSP as a good DM candidate. However, $M$ theory usually provides axion dark matter candidates.

In the end we have an $SO(10)$ model which has a tree-level, renormalisable, discrete preserving superpotential of the form
\beq
W = y_u  H_u \left( Q u^c +L N \right) + y_d H_d \left( Q d^c + L e^c \right)+ M \overline{d^c}_X d^c_X ,
\eeq
where family indices are understood, numerical coefficients omitted, and due to $SO(10)$ GUT symmetry $y_u = y_d$.

On top of this, there is an effective superpotential generated by Kahler interactions as the moduli aquire vevs. This superpotential is taken in the supergravity flat-limit and includes: $\mu$-terms generated \`a la Guidice-Masiero for the MSSM Higgses, the new vector-like family, and bilinear couplings between the extra matter and regular matter; RPV interactions that allows for the LSP to decay below current bounds but consequently cannot be a DM candidate. Furthermore, the full field theory has a universal soft-masses Lagrangian where the soft-masses' values are set by the gravitino mass.

\begin{figure}[h]
\begin{center}
\includegraphics[scale=1]{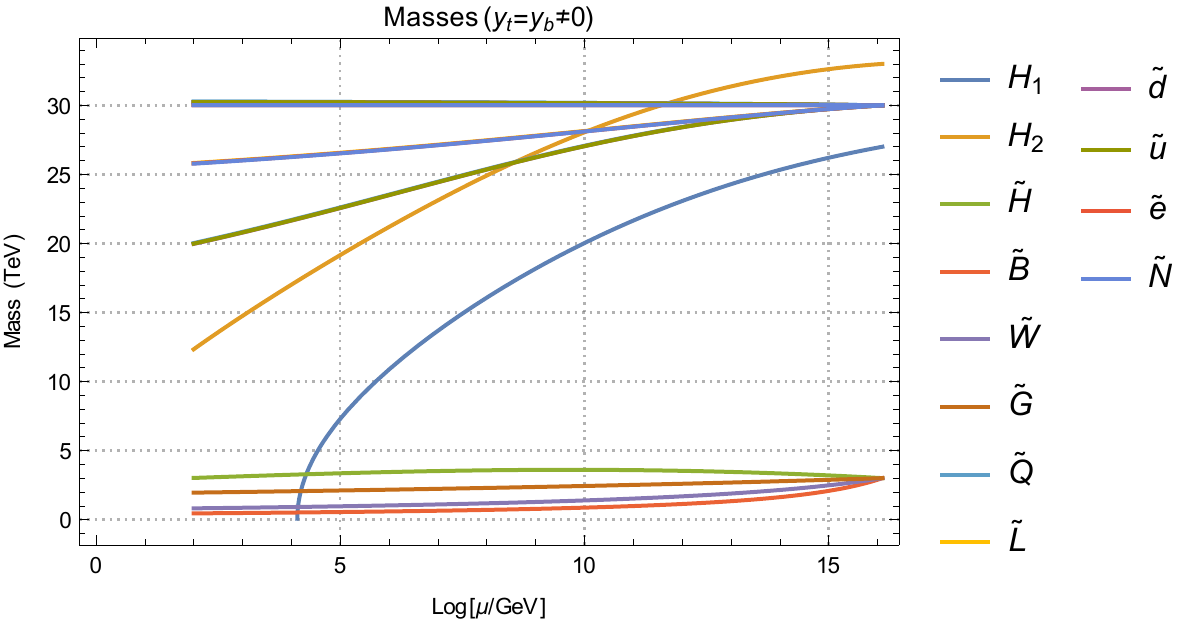} 
\caption{Illustrative example of the RGE analysis for the spectrum of an $SO(10)$ model}\label{fig:RGEmasses}
\end{center}
\end{figure}

The low energy spectrum can be studied through the regular RGE analysis. A very general outline of the expected spectrum can be seen in figure \ref{fig:RGEmasses}, where only the third family scalar partners are shown and we set $m_{3/2} = 30$ TeV, soft scalar masses at the Planck scale $m^2 =\left( m_{3/2}\right)^2$, $\mu$-terms and soft gaugino masses at the Planck scale $\mu = m_{1/2} = 0.1 m_{3/2}$. The running shows that radiative symmetry breaking for the EWS is generally easy to achieve within our framework.


\section{Conclusions}

We have discussed the origin of an $SO(10)$ SUSY GUT from $M$ theory on a $G_2$ manifold. The constructed model features striking predictions for the LHC as we are forced to include a light extra vector-like family ${\bf 16}_X+{\bf {\overline{16}}}_X$, which we were led to add by a novel solution for the doublet-triplet splitting problem. Proton-decay operators are forbidden by the geometrical properties, namely the discrete symmetries, of the extra dimensions by decoupling the colour triplets from matter. The addition of the extra states, being full GUT multiplets with the colour triplets as $d^c_X, \overline d^c_X$ are integrated out, do not spoil gauge coupling unification. Apart from providing a mechanism to save gauge coupling unification from the light colour triplets, the extra vector-like family plays a crucial role in breaking the extra $U(1)$ gauge group , while providing a most welcomed scenario to employ a see-saw mechanism to explain the light physical neutrino masses. We also studied the consequences of the moduli stabilisation as their vevs break the geometrical discrete symmetry, and we found that newly generated RPV interactions cannot induce dangerous proton-decay operators but let the LSP decay to quickly to be a good DM candidate. An outline of the expected spectrum of models built in this approach, shows that the LHC will provide the ultimate test as the vector-like family ${\bf 16}_X+{\bf {\overline{16}}}_X$ and the gauginos lie in range of its searches.

\ack

This work was funded by FCT under the grant SFRH/BD/84234/2012. The author would like to thank the collaboration with whom this work was done, Bobby S. Acharya, Krzysztof Bozek, Stephen F. King, and Chakrit Pongkitivanichkul; and to J. C. Rom\~ao for proofreading this manuscript. 

\section*{References}
\bibliographystyle{iopart-num}
\bibliography{CRISPIRM_ROMAO_miguel_DISCRETE2014}

\providecommand{\newblock}{}
\begin{thebibliography}{10}
\expandafter\ifx\csname url\endcsname\relax
  \def\url#1{{\tt #1}}\fi
\expandafter\ifx\csname urlprefix\endcsname\relax\def\urlprefix{URL }\fi
\providecommand{\eprint}[2][]{\url{#2}}

\bibitem{Aad:2012tfa}
Aad G {\em et~al.\/} (ATLAS Collaboration) 2012 {\em Phys.Lett.\/} {\bf B716}
  1--29 (\textit{Preprint} \eprint{1207.7214})

\bibitem{Chatrchyan:2012ufa}
Chatrchyan S {\em et~al.\/} (CMS Collaboration) 2012 {\em Phys.Lett.\/} {\bf
  B716} 30--61 (\textit{Preprint} \eprint{1207.7235})

\bibitem{CMS-PAS-HIG-14-009}
 2014 {Precise determination of the mass of the Higgs boson and studies of the
  compatibility of its couplings with the standard model} Tech. Rep.
  CMS-PAS-HIG-14-009 CERN Geneva

\bibitem{Aad:2014lra}
Aad G {\em et~al.\/} (ATLAS Collaboration) 2014 {\em JHEP\/} {\bf 1410} 24
  (\textit{Preprint} \eprint{1407.0600})

\bibitem{Chatrchyan:2013lya}
Chatrchyan S {\em et~al.\/} (CMS Collaboration) 2013 {\em Eur.Phys.J.\/} {\bf
  C73} 2568 (\textit{Preprint} \eprint{1303.2985})

\bibitem{Aad:2014nua}
Aad G {\em et~al.\/} (ATLAS Collaboration) 2014 {\em JHEP\/} {\bf 1404} 169
  (\textit{Preprint} \eprint{1402.7029})

\bibitem{Acharya:2007rc}
Acharya B~S, Bobkov K, Kane G~L, Kumar P and Shao J 2007 {\em Phys.Rev.\/} {\bf
  D76} 126010 (\textit{Preprint} \eprint{hep-th/0701034})

\bibitem{Acharya:2008zi}
Acharya B~S, Bobkov K, Kane G~L, Shao J and Kumar P 2008 {\em Phys.Rev.\/} {\bf
  D78} 065038 (\textit{Preprint} \eprint{0801.0478})

\bibitem{Acharya:2012tw}
Acharya B~S, Kane G and Kumar P 2012 {\em Int.J.Mod.Phys.\/} {\bf A27} 1230012
  (\textit{Preprint} \eprint{1204.2795})

\bibitem{Kachru:2003aw}
Kachru S, Kallosh R, Linde A~D and Trivedi S~P 2003 {\em Phys.Rev.\/} {\bf D68}
  046005 (\textit{Preprint} \eprint{hep-th/0301240})

\bibitem{Balasubramanian:2005zx}
Balasubramanian V, Berglund P, Conlon J~P and Quevedo F 2005 {\em JHEP\/} {\bf
  0503} 007 (\textit{Preprint} \eprint{hep-th/0502058})

\bibitem{Witten:1985xc}
Witten E 1985 {\em Nucl.Phys.\/} {\bf B258} 75

\bibitem{Mohapatra:2007vd}
Mohapatra R~N and Ratz M 2007 {\em Phys.Rev.\/} {\bf D76} 095003
  (\textit{Preprint} \eprint{0707.4070})

\bibitem{Lee:2010gv}
Lee H~M, Raby S, Ratz M, Ross G~G, Schieren R {\em et~al.\/} 2011 {\em
  Phys.Lett.\/} {\bf B694} 491--495 (\textit{Preprint} \eprint{1009.0905})

\bibitem{Acharya:2001gy}
Acharya B~S and Witten E 2001  (\textit{Preprint} \eprint{hep-th/0109152})

\bibitem{Acharya:2004qe}
Acharya B~S and Gukov S 2004 {\em Phys.Rept.\/} {\bf 392} 121--189
  (\textit{Preprint} \eprint{hep-th/0409191})

\bibitem{Witten:2001bf}
Witten E  (\textit{Preprint} \eprint{hep-ph/0201018})

\bibitem{Acharya:2015oea}
Acharya B~S, Bozek K, Crispim Rom\~ao M, King S~F and Pongkitivanichkul C 2015
  (\textit{Preprint} \eprint{1502.01727})

\bibitem{Acharya:2008hi}
Acharya B~S and Bobkov K 2010 {\em JHEP\/} {\bf 1009} 001 (\textit{Preprint}
  \eprint{0810.3285})

\bibitem{Giudice:1988yz}
Giudice G and Masiero A 1988 {\em Phys.Lett.\/} {\bf B206} 480--484

\bibitem{Brignole:1997dp}
Brignole A, Ibanez L~E and Munoz C 2010 {\em Adv.Ser.Direct.High Energy
  Phys.\/} {\bf 21} 244--268 (\textit{Preprint} \eprint{hep-ph/9707209})

\bibitem{Dvali:1995hp}
Dvali G 1996 {\em Phys.Lett.\/} {\bf B372} 113--120 (\textit{Preprint}
  \eprint{hep-ph/9511237})

\bibitem{Kilian:2006hh}
Kilian W and Reuter J 2006 {\em Phys.Lett.\/} {\bf B642} 81--84
  (\textit{Preprint} \eprint{hep-ph/0606277})

\bibitem{Reuter:2007eh}
Reuter J 2007  (\textit{Preprint} \eprint{0709.4202})

\bibitem{Howl:2007zi}
Howl R and King S 2008 {\em JHEP\/} {\bf 0801} 030 (\textit{Preprint}
  \eprint{0708.1451})

\bibitem{Acharya:2014vha}
Acharya B~S, Kane G~L, Kumar P, Lu R and Zheng B 2014 {\em JHEP\/} {\bf 1410} 1
  (\textit{Preprint} \eprint{1403.4948})

\bibitem{Banks:1995by}
Banks T, Grossman Y, Nardi E and Nir Y 1995 {\em Phys.Rev.\/} {\bf D52}
  5319--5325 (\textit{Preprint} \eprint{hep-ph/9505248})

\bibitem{Hirsch:2000ef}
Hirsch M, Diaz M, Porod W, Romao J and Valle J 2000 {\em Phys.Rev.\/} {\bf D62}
  113008 (\textit{Preprint} \eprint{hep-ph/0004115})

\bibitem{Acharya:2011te}
Acharya B~S, Kane G, Kuflik E and Lu R 2011 {\em JHEP\/} {\bf 1105} 033
  (\textit{Preprint} \eprint{1102.0556})

\bibitem{Dreiner:1997uz}
Dreiner H~K 2010 {\em Adv.Ser.Direct.High Energy Phys.\/} {\bf 21} 565--583
  (\textit{Preprint} \eprint{hep-ph/9707435})

\end{thebibliography}

\end{document}